\documentclass[aps,prb,twocolumn,superscriptaddress,longbibliography,showkeys]{revtex4-2}

\usepackage[T1]{fontenc}
\usepackage[utf8]{inputenc}

\usepackage{amsmath}
\usepackage{amssymb}
\usepackage{amsfonts}
\usepackage{bm}
\usepackage{braket}
\usepackage{graphicx}
\usepackage{hyperref}
\usepackage{xcolor}
\usepackage{changes}

\newcommand{\ta}{t_{\mathrm A}}
\newcommand{\tb}{t_{\mathrm B}}

\begin{document}

\title{
Inversion-symmetric topological insulators in cut-and-project binary chains
}

\author{Zhipeng Zeng}
\thanks{Co-first author} 
\affiliation{Institute for Quantum Science and Technology, Shanghai University, Shanghai 200444, China}
\author{Yuge Chen}
\thanks{Co-first author} 
\email{Contact author : chenyuge@shu.edu.cn}
\affiliation{Institute for Quantum Science and Technology, Shanghai University, Shanghai 200444, China}
\author{Jean-Noël Fuchs}
\email{Contact author : jean-noel.fuchs@sorbonne-universite.fr}
\affiliation {Sorbonne Université, CNRS, Laboratoire de Physique Théorique de la Matière Condensée, LPTMC, F-75005 Paris, France}
\author{Jianxin Zhong}
\email{Contact author : jxzhong@shu.edu.cn}
\affiliation{Institute for Quantum Science and Technology, Shanghai University, Shanghai 200444, China}
\author{Rémy Mosseri}
\email{Contact author : remy.mosseri@sorbonne-universite.fr}
\affiliation{Sorbonne Université, CNRS, Laboratoire de Physique Théorique de la Matière Condensée, LPTMC, F-75005 Paris, France}

\date{\today}


\begin{abstract}

 We investigate the electronic properties of binary tight-binding chains generated by the cut-and-project method for rational slopes $\alpha=p/q$, leading to periodic and inversion symmetric chains with $n=p+q$ sites. The binary structure is encoded in two hopping amplitudes $\ta$ and $\tb$. For fixed $\ta \neq \tb$, the support of the energy spectrum as a function of $p/n$ gives rise to a “Cut-and-Project butterfly”. We concentrate on insulators with $M$ filled bands among a total of $n$ bands and vary $\ta/\tb$. Inversion symmetry constrains the electric polarization $P$ to $0$ or $P_q/2$ modulo a polarization quantum $P_q = \gcd(M,n)/n$. A topological transition, between two insulators that differ by their quantized polarization, occurs if and only if $n/\gcd(M,n)$ is odd. When $n/\gcd(M,n)$ is even, the two insulating regimes have a vanishing polarization and no topological transition occurs,
  despite the gap closing at $\ta =\tb$. When $n$ is even and $M$ odd, we find an adiabatic path between $\ta>\tb$ and $\ta<\tb$ that maintains inversion symmetry and a gap.
 \end{abstract}


\maketitle

\section{Introduction}

The discovery of topological phases of matter has profoundly modified the
conceptual framework used to classify condensed-matter systems \cite{Thoules_1982,Haldane_1988,Kane_2005,Fu_2007,Cayssol_2021}. Beyond the
traditional paradigm based on symmetry breaking and local order parameters,
some topological phases are characterized by global properties of the electronic
wavefunctions that remain invariant under continuous deformations of the
Hamiltonian. Over the last decades, this perspective has led to the
identification of numerous classes of topological insulators and
superconductors, whose classification is strongly influenced by the symmetries
of the underlying system \cite{Altland_1997,kitaev2009periodic,Ryu_2010}.

Among crystalline symmetries, inversion symmetry occupies a particularly
important position. Unlike time-reversal or particle-hole symmetries,
inversion acts directly in real space and imposes strong constraints on the
electronic structure. In one-dimensional insulating systems, inversion
symmetry quantizes geometric quantities such as the Zak phase and restricts
the possible locations of Wannier centers. This leads to strong constraints on measurable quantities such as the electric polarization, as discussed in the so-called 
`` modern theory of electric polarization" developed by
King-Smith, Vanderbilt and Resta \cite{Zak1989,KingSmith1993,Vanderbilt_1993,Resta1994}.

One-dimensional systems have long served as valuable theoretical laboratories
for the study of topological phenomena \cite{Hasan2010}. The Su-Schrieffer-Heeger (SSH) model \cite{SSH1979},  among others, is often regarded as a paradigmatic example of a one-dimensional topological insulator, owing to its simple structure and the existence of localized boundary mid-gap states. 

More recently, however, it has been emphasized that the
interpretation of topological phase transitions (TPT) in one dimension requires particular
care, as displayed for instance when computing the electric polarization \cite{FuchsPiechon2021}. Key concepts such as the existence of a quantum of polarization, and the natural requirement that physical results should not depend on a specific choice of the unit cell location in a periodic system, are needed in order to provide solid results and discard ambiguities.

The purpose of the present work is to investigate, along these lines, a broad family of
binary tight-binding chains some of which exhibit topological phase transitions. These structures are  generated by connected sequences of vertical ($A$ type) and horizontal ($B$ type) edges selected from a square lattice, close to a line of rational slope, and are inversion symmetric by construction. 
The selection method is that given  by the so-called cut-and-projection (CP) method, which was originally introduced in the context of quasicrystals and their periodic approximants ~\cite{Duneau1985,KKL1985, Elser1986}. 
In one dimension, rational cuts generate periodic binary sequences, while irrational cuts lead to
quasiperiodic structures.

In the present construction, the binary $(A,B)$ sequences
will be those given by  the ``cut" part of the CP method, along a line of rational slope $\alpha=p/q$, the resulting structure being periodic with $n=p+q$ sites per unit cell.
 In the standard CP algorithm, this staircase would be then projected onto the line of slope $\alpha$, leading to a 1d chain with two different bond lengths. Here, for the sake of simplicity, the whole selected structure is projected onto a line of slope unity, leading to equally spaced points separated by the selected periodic sequence  of $A$ and $B$ bonds.  

The systems considered here are described by a nearest-neighbor tight-binding Hamiltonian for spinless electrons with a single orbital per site,

\begin{equation}
H=-\sum_i t_{i,i+1} \ket{i} \bra{i+1} + t_{i+1,i} \ket{i+1} \bra{i}
\label{eq:hamiltonian}
\end{equation}

where the off-diagonal hopping amplitudes belong to the binary set $t_{i,j}\in\{\ta,\tb\}$.

The family of chains considered in this work possesses several remarkable
properties. First, all structures are inversion symmetric and therefore admit
inversion centers that constrain the possible Wannier-center locations.
Second, exchanging the two bond types $A$ and $B$ maps the sequence
associated with a slope $\alpha$ onto the sequence associated with the dual
slope $1/\alpha$. Within the entire family, the slope $\alpha=1$ is special, as discussed in Sec.~\ref{sec:duality}.

Another important observation concerns the role of the hopping ratio
$\ta/\tb$. When $\ta=\tb$, all binary structures become equivalent to a
uniform periodic chain and the spectral gaps close. Away from this point,
gaps open and insulating phases can emerge for specific electron densities. As will be shown below, these phases
can be characterized through the combined analysis of Wannier centers,
electric polarization and inversion symmetry. We show that the occurrence of a phase transition across the $\ta=\tb$ value depends on a simple arithmetical criterion involving the period $n$ and the number of occupied bands $M$, with $M \in \{1,n-1\}$.

The paper is organized as follows. In Sec.~\ref{sec:cut_projection} we review
the construction of inversion-symmetric binary chains from rational
CP and discuss the associated duality between slopes
$\alpha$ and $1/\alpha$. Section~\ref{sec:spectrum} introduces the tight-binding
Hamiltonian and analyzes its general spectral properties. In
Sec.~\ref{sec:farey} we present a butterfly-like diagram that summarizes the
global organization of the spectra as a function of the rescaled slope 
$\theta= p/n= \alpha/(1+\alpha) $.
Sections~\ref{sec:oddn} and \ref{sec:evenn} are devoted respectively to 
simple $n$ odd or even representatives of the family. These two examples reveal the essential mechanisms governing the
evolution of Wannier centers and polarization across the gap-closing point
$\ta=\tb$. Section~\ref{sec:general} extends the discussion to arbitrary
unit-cell sizes and proposes a general classification of the insulating
phases. Open boundary conditions (OBC) are discussed in Sec.~\ref{sec:edges}.
Sec.~\ref{sec:conclusion} summarizes the main results and outlines several
possible extensions of the present work.
An analysis of the CP butterfly skeleton is given in appendix~\ref{app:farey discussion}. Appendix~\ref{app:even n case} presents an important construction showing the absence of TPT for even $n$ and odd $M$. A peculiar form of edge states, for $\alpha=1/q$, is described in appendix~\ref{app:ABn}. Finally, appendix~\ref{app:diagonal term} shows that introducing a diagonal term in the Hamiltonian preserving inversion symmetry follows our general result on the occurrence of TPTs.


\section{Inversion-symmetric binary chains from the cut-and-projection method}
\label{sec:cut_projection}

\subsection{Rational cut-and-projection construction}

\begin{figure}[ht]
    \centering
    \includegraphics[width=\columnwidth] {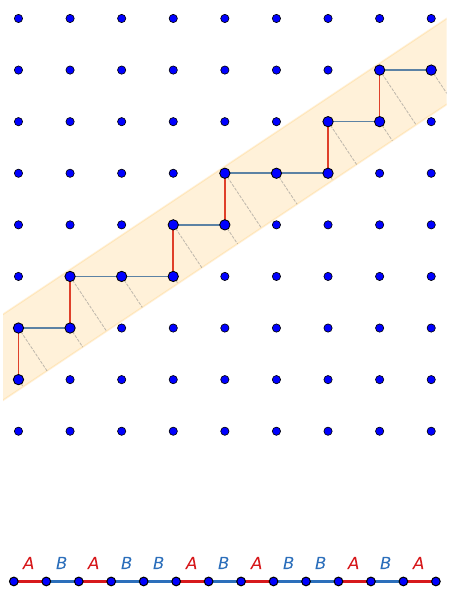}
       \caption{
        Schematic of the cut-and-projection construction.Top: The yellow strip region (the "acceptance zone) with width $W$ and slope $\alpha=2/3$ selects points in the $\mathbb{Z}^2$ lattice, and edges connecting these points (noted $A$ for the vertical edges and $B$ for horizontal edges).
      This set of selected points is usually projected onto the physical space
       $E_\parallel$ along the dotted lines, leading to  projected edges of different length. Bottom: In the present work, the selected points in the strip are instead projected onto a line of slope unity, leading to $A,B$ bonds of equal length.}
     
    \label{fig:CP}
\end{figure}

The binary bond sequences considered in this work are generated by a
CP construction from the square lattice
$\mathbb{Z}^2$.
One considers a particular line of slope $\alpha$, routinely called the ``parallel" or ``physical" space, noted $E_\parallel$. An ``acceptance" semi-open strip is formed by translating the  $\mathbb{Z}^2$ unit square along this line. All sites located inside this strip are selected, as well as the edges connecting  selected sites,  forming a staircase-like structure, noted $\Sigma_\alpha$, with an $\alpha$-dependent succession of vertical (type $A$) and horizontal (type $B$) edges (see Figure~\ref{fig:CP}).  

An equivalent way to select the sites is to project the $\mathbb{Z}^2$ sites onto the line perpendicular to $E_\parallel$, usually called the ``perp" space direction $E_\perp$. Only those sites which are mapped inside a semi-open interval of precise length will be selected. This interval corresponds to the shadow of the $\mathbb{Z}^2$ unit square projected onto $E_\perp$, and has width  $W=(p+q)/\sqrt{(p^2+q^2)}$. If the slope $\alpha$ is irrational, the selected sequence  is quasiperiodic. The standard Fibonacci chain is constructed this way, upon mapping the selected staircase onto the physical line,  whose slope is related to the golden mean. In the context of quasiperiodic structures, the $E_\perp$ coordinate is also called a phason coordinate.

In the present work, we consider straight lines of rational slope $\alpha=p/q$, limiting for the moment to $\; 0<\alpha\le 1$, where $p$ and $q$ are coprime integers.

This defines a binary word of length $n$,

\begin{equation}
S(\alpha)=(s_1,s_2,\ldots,s_n),
\qquad
s_j\in{A,B},
\end{equation}

which determines the unit cell hopping amplitudes of the tight-binding model
introduced in the next section.
For a rational slope, the infinite periodic chain corresponds to an infinite repetition of the $S(\alpha)$ sequence.

Typical $S(\alpha)$ examples are

\begin{align}
\alpha &=1
\rightarrow
AB ,
\
\alpha =\frac12
\rightarrow
ABB ,
\
\alpha =\frac13
\rightarrow
ABBB \nonumber,
\\
\alpha &=\frac23
\rightarrow
ABBAB .
\end{align}

An important consequence of the restriction $\alpha\le 1$ is that the sequence never contains two consecutive $A$ bonds. Notice that, for fixed rational $\alpha$, the $S(\alpha)$ sequence is unique up to circular permutation, corresponding to the possible different choices of the unit cell in a periodic structure.

\subsection{Projected chain and physical space}

In the standard CP method, the selected sites $\Sigma_\alpha$ are projected onto $E_\parallel$, leading to a 1D chain with two unequal edge lengths, corresponding to the different projections of  vertical and horizontal  segments $A$ and $B$. However, in the CP method, one can decouple the selection step from the projection, and choose a physical space different from the initial line of slope $\alpha$. The resulting sequence of bonds (which contains a large part of the chain complexity) will be identical, but the metric part will differ.  For the purpose of the present work, we find it convenient to project the staircase structure $\Sigma_\alpha$ onto the line of slope unity. After projection, all sites become equally spaced along this new physical
direction. The position of the $n$ sites inside the projected unit cell can
therefore be chosen as

\begin{equation}
x_i=
\frac{i-1}{n}a,
\qquad
i=1,\ldots,n ,
\end{equation}

where $a$ denotes the unit cell length. In the following, we will use units such that $a=1$.

As a consequence, the geometrical complexity of the approximant is no
longer encoded in the site positions but entirely in the binary bond
sequence itself.

This choice  simplifies the construction of the canonical Bloch
Hamiltonian and the calculation of geometrical quantities such as
Zak phases, Wannier centers and electric polarization but should not affect the topological phase classification that will be presented later.

\subsection{Duality under bond exchange}
\label{sec:duality}

Exchanging the labels $A$ and $B$ has a simple geometrical
interpretation.
Within the square-lattice construction, this operation amounts to
interchanging the horizontal and vertical directions. Therefore, a
chain generated from a slope $\alpha$ is mapped onto the chain
generated from the inverse  slope

\begin{equation}
\alpha
\longrightarrow
\frac{1}{\alpha}.
\end{equation}

For this reason, it is sufficient to restrict the discussion to the
interval $0<\alpha\le 1$, 
the complementary interval behaviour being derived upon exchanging the two hopping
amplitudes
$t_A
\longleftrightarrow
t_B.$

A remarkable exception occurs for $\alpha=1$,
which is the fixed point of the transformation. The corresponding (SSH) sequence
$ABABAB\ldots$ 
is invariant under the exchange of the two bond types, up to a
translation by one bond.

A last remark is in order here. The duality transformation is not just a nice geometrical symmetry which simplifies the general analysis by restricting it to the $\alpha<1$ case. It will also play an important role in our treatment of the even $n$ case, as discussed later (see Sec.~\ref{sec:general even}).

\subsection{Inversion symmetry}
\label{sec:inversion}

The one-dimensional periodic chains constructed by the CP method possess strict space-inversion symmetry at any rational slope, a property jointly guaranteed by the inversion symmetries of the initial two-dimensional square lattice $\mathbb{Z}^2$ and those which remain in the projection strip $\Sigma_\alpha$.

The  $\mathbb{Z}^2$ lattice has three types of inversion centers, either at sites, mid-edges and square centers. We look for inversions which are preserved in the selected strip. Due to the semi-open nature of the strip, we must discard $\mathbb{Z}^2$ inversion centers located at  square centers, since two opposite sites on a $\mathbb{Z}^2$ square cannot be both selected.

Now consider the projection of the strip-selected sites onto $E_\perp$. For a rational cut, sites project onto $n$ equally spaced positions, a property that was used in the ``conumbering" approach to quasiperiodic approximants \cite{Mosseri1988,SireMosseri1989}, which amounts to numbering the sites inside a unit cell according to their (periodically spaced) coordinate in $E_\perp$. We can now shift the whole strip along $E_\perp$ such that its central line goes through a $\mathbb{Z}^2$ inversion symmetry center, at which we place the origin of the coordinate system, which we now define, after proper rotation, in an orthonormal basis of $(E_\parallel,E_\perp)$. Whenever $n$ is even, the central line can only go through a mid-edge, while for $n$ odd, it can also meet a site.

With this strip position, for any site inside the strip, with coordinate ($x_\parallel,x_\perp)$ its image under this inversion center reads ($-x_\parallel,-x_\perp)$ also belongs to the strip, showing that $\Sigma_\alpha$ is inversion symmetric. This property holds after projection onto the physical space (here, as said above, the line of slope unity), which proves that all periodic chains considered here possess inversion symmetry.

Now, a well known property of a 1D inversion symmetric periodic structure is that each unit cell contains a second inversion center, shifted by half a unit-cell length. This leads to a distinction between odd and even chains:  with odd $n$, this pair of inversion centers are of type (site/mid-edge), while for even $n$ it is of type (mid-edge $A$/mid-edge $B$).

With our coordinate system choice, all $\Sigma_\alpha$ inversion centers project onto the $E_\perp$ origin. This explains the absence of inversion center pairs of type (site/site): since $p$ and $q$ are coprime, the shortest distance between two sites in the selected strip is equal to the unit cell length, and therefore not half this value. This also applies to mid-edge inversion center pairs, which must be of different type ($A$ and $B$).

As will be shown below, the positions of these inversion centers play a fundamental role in constraining the possible Wannier-center locations and therefore the topological properties of the insulating phases. In particular, they naturally lead to different behaviors for odd and even unit cells, a distinction that will be made more precise below.


\section{Tight-binding model on Cut and Project chains}
\label{sec:spectrum}

\subsection{Canonical Bloch Hamiltonian}
\label{canonical}

Using Bloch's theorem, the Hamiltonian given in Eq.~\ref{eq:hamiltonian} is transformed into a set of $k$-dependent $n\times n$ Hamiltonians written in a canonical Bloch basis :

\begin{equation}
|k,j\rangle
=
\frac{1}{\sqrt{N}}
\sum_m
e^{ik(m+x_j)}
|m,j\rangle .
\end{equation}
with $m\in \mathbb{Z}$ labelling the different periodic cells, and $j$ the sites in a unit cell.
In this basis, all nearest-neighbor hoppings acquire the same geometric
phase factor and can be written as

\begin{equation}
T_j(k)=t_j\,e^{ik/n},
\qquad
j=1,\ldots,n,
\label{eq:Tj}
\end{equation}

where $t_j\in\{\ta,\tb\}$ follows the binary bond sequence.

The Bloch Hamiltonian takes the cyclic tridiagonal form

\begin{equation}
H(k)=
\begin{pmatrix}
0 & T_1 & 0 & \cdots & T_n^{*} \\
T_1^{*} & 0 & T_2 & \cdots & 0 \\
0 & T_2^{*} & 0 & \cdots & 0 \\
\vdots & & & \ddots & T_{n-1} \\
T_n & 0 & 0 & T_{n-1}^{*} & 0
\end{pmatrix},
\label{eq:Hgeneral}
\end{equation}

where the $k$-dependence is entirely contained in the coefficients defined in Eq.~(\ref{eq:Tj}).


At $\ta=\tb$, the Hamiltonian reduces to that of a uniform periodic chain. The binary modulation disappears and the gaps induced by the enlarged unit cell close. For $\ta\neq\tb$, the modulation opens gaps and insulating phases can occur when the Fermi energy lies inside one of them. For any periodic structure with $n$ sites per unit cell, the spectrum consists of $n$ continuous bands generically separated by $n-1$ gaps, and the eigenfunctions are Bloch waves.

Electronic properties of quasiperiodic  1D chains have been extensively studied since the pioneering work of Kohmoto et al.~\cite{Kohmoto_1983} and Ostlund et al ~\cite{Ostlund_1983}, with a growing interest after the discovery of quasicrystals~\cite{Shechtman_1984} , the quasiperiodic Fibonacci chain being the simplest case of such a system, displaying an unusual spectrum. The latter was studied along different methods, taking advantage of the self-similar/hierarchical properties of this chain, which are nicely expressed by its successive periodic approximants. Topological properties have also been investigated (see for instance Ref.~\cite{Naumis2008, Kraus2012, Baboux2017, Kellendonk2019,jagannathan2025}). 
We do not intend to summarize here the large set of results obtained for the Fibonacci chain (and many other chains constructed with CP method applied to slopes given by quadratic irrational numbers). Let us just recall that their spectrum has singular measure, and their eigenfunctions are generically critical (neither localized nor extended).  

 The quasiperiodic chain properties can be approached step by step upon studying the sequence of best rational approximation of the associated irrational numbers. The transition from continuous bands/Bloch states to singular continuous bands/critical states only appears after a careful analysis of the spectra of the successive  best approximant structure.

Here, we aim to study the topological properties (if any) of the rational periodic structures studied for themselves, not in terms of their potential use to address the quasiperiodic case (which is let for future work). Since these topological features are expected for insulator phases with Fermi level lying in the gaps, let us first consider the set of bands and gaps for the different rational periodic chains.

\subsection{CP butterfly}
\label{sec:farey}

\begin{figure}[t]
\centering
 \includegraphics[width=\columnwidth]{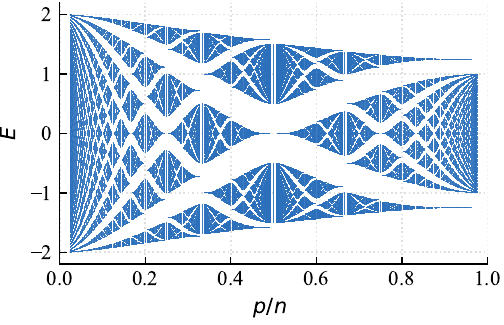}
\caption{
CP butterfly for the binary tight-binding chains. For each rational
slope $\alpha=p/q$, the unit cell contains $n=p+q$ sites and the spectrum
contains $n$ bands. The spectrum support is plotted against the rescaled CP slope $\theta=p/n$, for hopping amplitudes $\tb =1$ and $\ta=0.5$. This diagram displays a rich hierarchical structure.
}
\label{fig:farey}
\end{figure}

 Figure~\ref{fig:farey} displays the support of the spectrum for the different rational periodic structures, computed in the case $\tb=2 \ta$, and here given as a function of the rescaled slope $\theta=\alpha/(1+\alpha) = p/n$.

The spectra are computed numerically for all rational numbers with denominators smaller or equal to a maximal value $n_{max}$, a so-called ``Farey sequence", and display a nice intricate gap pattern associated with the particular structure of the CP sequences; this is why we propose to call this pattern the ``CP butterfly", in contrast  to the celebrated Hofstadter butterfly \cite{Hofstadter_1976} associated with tight-binding electrons under perpendicular magnetic field in $\mathbb{Z}^2$. The rational number refers to the magnetic flux (in unit of the  quantum flux) in the latter case, while it refers here to the rescaled slope $\theta$.

To the best of our knowledge, a similar butterfly, but corresponding to a diagonal modulation of the Hamiltonian, was first displayed by Ostlund and Kim in 1985~\cite{OstlundKim1985}, when extending to arbitrary slopes previous analyses focused on the golden-mean case~\cite{Kohmoto_1983,Ostlund_1983}; see note~\cite{terminology}.  

The Farey hierarchy naturally governs the multiscale structure of rational CP chains.
Let $ S\left(p_1/q_1\right)$ and $S\left(p_2/q_2\right) $ be the $(A,B)$ words associated with rational slopes $p_1/q_1$ and $p_2/q_2$ which are neighbouring rationals in a given Farey series. In a higher order Farey series, the mediant $(p_1+p_2)/(q_1+q_2)$ will eventually appear whose associated $(A,B)$ word reads

$$  S\left(\frac{p_1+p_2}{q_1+q_2}\right)= S\left(\frac{p_1}{q_1}\right)\oplus S\left(\frac{p_2}{q_2}\right), $$

where $\oplus$ denotes word concatenation. This property acquires a direct geometrical interpretation: it expresses how large rational CP chain unit cells can be decomposed into more elementary ones organized according to the Farey tree. This allows for a simple numerical computation of the CP butterfly in terms of hierarchical products of transfer matrices, as was used in Ref.~\cite{OstlundKim1985}.

A simple way to analyze the structure of this CP butterfly is to consider what happens in perturbation for $\ta$ close but different from $\tb$, see ref \cite{SireMosseri1990}. This is described in appendix~\ref{app:farey discussion}.

\subsection{Zak phase and Wannier centers}
\begin{figure}[t]
\centering
 \includegraphics[width=\columnwidth]{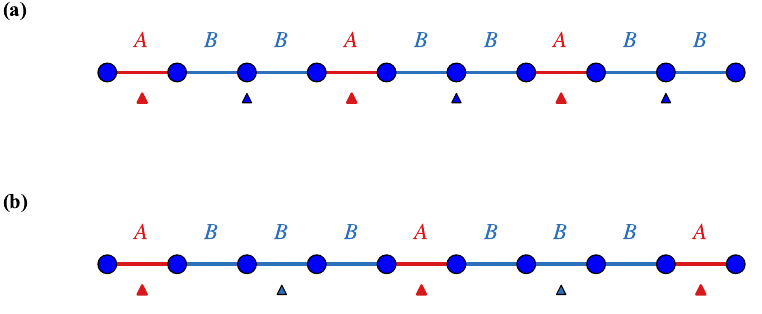}
\caption{
Wannier-center configurations for odd and even chains. (a) $\alpha=1/2$ ABB chain: Wanniers centers sit either at a site (shared by two B bonds) or at a bond center (here an A bond); (b) $\alpha=1/3$ ABBB chain: the  Wannier centers remain bond-centered in both insulating
regimes, but occupy different inversion-symmetric positions (here a A bond or a B bond centers).
}
\label{fig:wannier_center}
\end{figure}

For a one-dimensional inversion-symmetric insulator, the topological character of its bands is captured by the Zak phase\cite{Zak1989}, which is simply related to the position of the Wannier center in real space\cite{KingSmith1993,Resta1994}. Whenever inversion symmetry is present, the latter is constrained to sit at one of the inversion centers.  This is  shown in Fig~\ref{fig:wannier_center}, in the case $(p,q)=(1,2)$ and $(1,3)$.

The Zak phase of the $m$‑th isolated band is defined as
\begin{equation}
    Z_m = \int_{-\pi}^{\pi} dk \langle u_{m,k} \mid i\partial_k u_{m,k} \rangle 
    + arg \langle u_m(-\pi) \mid e^{i2\pi x} \mid u_m(\pi) \rangle,
    \label{eq:zak}
\end{equation}
where $\lvert u_{m,k}\rangle = e^{-ikx} \lvert \psi_{m,k}\rangle$ is the cell‑periodic part of the Bloch function and $x$ is the position operator~\cite{Resta_1998}. The second term on the right side  stems from the fact that $\lvert u_{mk}\rangle$ is generally not periodic across the Brillouin zone boundary.
The bare Berry‑connection integral therefore does not run over a closed loop, and its value depends on the choice of gauge. The second term closes the path explicitly by inserting the matrix element of $e^{i2\pi x}$ between the two boundary states, thereby removing the gauge ambiguity and establishing a direct relation to the Wannier center: $Z_m = 2\pi \langle x_m \rangle$~\cite{Vanderbilt_1993}.

In the numerical calculations, we use standard Born-von Karman periodic boundary conditions, closing the chain onto itself after $L$ repeated unit cells. As a result, reciprocal space is discretized, with $L$ points in the first Brillouin zone. As usual, the integral in Eq. \ref{eq:zak} is transformed into a Wilson loop, via the computation of $\exp \mathrm{i} Z_m$ still taking care of the above discussed closing term.

An important point here concerns the precise definition of the position operator. As is well known, the standard position operator $x$ cannot be used with periodic boundary conditions, and is replaced, along the Resta prescription \cite{Resta_1998}, by a unitary operator which sends the position to a phase information measured on the closed chain containing $L \times n$ relative phases. Fixing the phase origin, we can keep track of the position in the closed chain, locate the Wannier centers for each band, and verify that they are located at the chains inversion centers (repeated inside each unit cell).

A given insulating rational periodic chain is then fully characterized by the computed set of its Wannier centers (one per filled band), which stays constant as long as gaps do not close (for $\ta=\tb)$.  As discussed above in section \ref{sec:inversion}, the possible inversion centers are of two types (see Fig.~\ref{fig:wannier_center}), depending on the parity of $n$.

\subsection{\label{sec:indicator}1D Inversion Symmetry Indicator}

In systems with spatial inversion symmetry, the computation of the Zak phase can be greatly simplified via the use of symmetry indicators\cite{Bernevig_2013}. The latter reduce the Berry-connection integral over the entire Brillouin zone to a product of parity eigenvalues of the occupied states at a few high‑symmetry points . For a one‑dimensional periodic structure, there are two such points at $k=0$ and $k=\pi$.

Denoting the parity eigenvalue of the $m$‑th occupied band at $k$ by $\xi_m(k)=\pm1$, the Zak phase of the system is determined by
\begin{equation}
    (-1)^{Z/\pi} = \prod_{k=0,\pi} \prod_{m \in m_{\mathrm{occ}}} \xi_m(k),
\label{eq:indicator}
\end{equation}
where $Z=0$ for a product of $+1$ on the RHS and $Z=\pi \pmod{2\pi}$ for $-1$.
In addition to the presence of inversion symmetry, this criterion relies on two prerequisites:  (i) the occupied bands must be isolated from one another between the high symmetry  points, i.e., there should be no band crossings or degeneracies; (ii) the origin of the coordinate system is chosen at an inversion center.
If the origin is shifted away from an inversion center, an extra phase factor depending on the shift appears in the formula.

In this work, we use the symmetry indicator given in Eq.~\ref{eq:indicator} to compute the Zak phases.
 Each filled band carries its own contribution to this product (the Zak phase associated to this band). We aim to compare, for a given insulating phase,  the total Zak phase changes across the critical point $\ta=\tb$. 
 
 With our choice of a global negative term in the Hamiltonian, the ground state in the lowest band occurring at $k=0$ is symmetrical (bonding state), contributing a factor $+1$ to the indicator. As a result, the possible change in the  Zak phase for this lower band will just be given by a symmetry change of the state at the boundary  of the Brillouin zone ($k= \pi$).  

We now discuss the important result, that the only contributions that can change across the critical point are those given by the two extremal bands, and therefore only the lowest band for a Fermi level in one of the $n-1$ gaps. The lowest band behaviour therefore dictates the Zak phase possible change.

Let us give an explanation of this statement within a perturbative picture. Consider an effective two‑level Hamiltonian for a pair of adjacent bands near a high‑symmetry point:
\begin{equation}
 H_{\mathrm{eff}} = 
 \begin{pmatrix}
     \epsilon & \delta\\
     \delta & \epsilon
 \end{pmatrix},
\end{equation}
where the hopping imbalance $\delta$ measures the deviation of the hoppings terms $(\ta,\tb)$ from the critical points. We write $t_A = 1+\delta$ and $t_B=1-\delta$.
At $\delta = 0 $ the two bands are degenerate at the high-symmetry point, and the gap closes (at energy  $\epsilon$).
When crossing the critical point, the upper and lower eigenstates of the above $H_{\mathrm{eff}}$ exchange their respective parity.

\begin{figure}[ht]
    \centering
      \includegraphics[width=\linewidth]{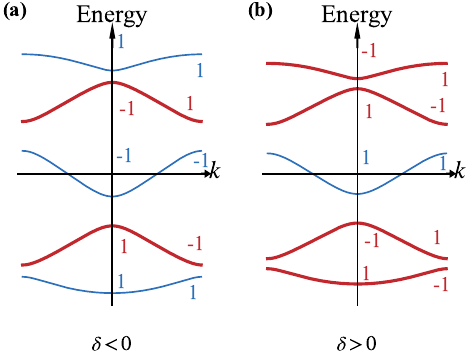}
    \caption{Band structures for the chain corresponding to $p/q=2/3$. The hopping amplitudes are $t_A=1+\delta$ and $t_B=1-\delta$ (a) $\delta<0$ and (b) $\delta>0$. The numbers $\pm1$ at the high‑symmetry points $k=0$ and $k=\pi$ label the inversion eigenvalues of each band. For bands coloured red, the product of inversion eigenvalues is $\xi_{k=0} \xi_{k=\pi} =-1$, corresponding to a Zak phase of $\pi$; for bands coloured blue, the product is $+1$,corresponding to a Zak phase of $0$. The gaps close at $\delta =0$.}
    \label{fig:3}
\end{figure}

 The lowest band only faces a  gap closing at its upper part (here for $k=\pi$).
Consequently, it always experiences a single parity inversion. This is also the case for the uppermost band.
In marked contrast, any intermediate band interacts with both its lower and upper neighbors, with a parity flip for each of the two avoided crossings, which therefore cancels out(see Fig.~\ref{fig:3}).

As a consequence, for any insulating phase (i.e. at any integer band filling), the symmetry indicator always displays a sign change at the critical point, and therefore a $\pi$ shift for the Zak phase. This is however not a sufficient condition for indicating a topological phase transition, as already noted in the SSH case~\citep{FuchsPiechon2021}. We must now compute the polarization to better qualify the two phases across the critical point.

\subsection{Polarization}

The potential presence of topological phase transitions in a one-dimensional inversion-symmetric insulator can be read out  on a measurable quantity, the electrical polarization, which contains both electronic and ionic parts. 
One follows the ``localization prescription" which consists in replacing the extended electronic charge distribution by point charges located at the Wannier centers when computing the electronic contribution to the polarization. The latter is directly related to the total Zak phase $Z$ of the occupied subspace~\cite{Zak1989,KingSmith1993}. To simplify notations, we fix from now on the electronic charge to $-1$; the electronic part of the polarization then reads:

\begin{equation}
    P_{\mathrm{elec}} = - \frac{1}{2\pi}Z \pmod{1}.
    \label{eq:Pelec}
\end{equation}
The $\mod{1}$ expresses the fact that there is one  Wannier center per unit cell.

Computing the electric polarization requires a neutral system, leading therefore to an ionic contribution (i.e. each site carries an equal positive charge). In addition, in a periodic system, a uniquely defined quantity requires to mod out the computed quantities by a ``quantum of polarization" $P_q$,  which characterizes the effect of equivalent  unit cell  definitions ~\cite{KingSmith1993,Spaldin_2012}, so that the total electric polarization reads: 
\begin{equation}
P=P_{elec}+ P_{ion} \, \pmod{P_q}.
\end{equation}

This $\mod{P_q}$ expresses the fact that in a periodic solid, polarization is not a simple scalar value, but 
is defined modulo a lattice with period $P_q$. For our CP chains with $n$ sites per unit cells, and $M$ filled bands, the neutrality condition is fulfilled by assigning a positive charge of $M/n$ to each site, so that the ionic contribution to the polarization reads:
\begin{equation}
P_{ion} = \sum_{j} q_j x_j = M/n \sum_{j} x_j,
\label{eq:Pion}
\end{equation}
where $j$ runs in the unit cell.
Shifting by one site the unit cell leads to a shift of $M/n$ for $P_{ion}$. Together with the modulo 1 operation in the computation of $P_{elec}$  (see Eq.\ref{eq:Pelec}, we now have that the polarization quantum  minimizes the quantity
$$
 k_1 + k_2 M/n, \quad  (k_1,k_2)\in \mathbb{Z}^2,
$$ 
which leads to 

\begin{equation}
P_q=\gcd{(M,n)}/n,
\label{eq:Pq}
\end{equation}

as was already derived in Ref.~\cite{Ayashi2021}.

In addition, inversion symmetry further simplifies the analysis. Wannier centers are now frozen  at one of the two inversion-center positions (see Sec.~\ref{sec:inversion}), and therefore  only take two possible values (here $0 , 1/2 \mod{1}$, by fixing the coordinates origin  at one of the inversion center). To further simplify the analysis, we take a unit cell such that the coordinate origin sits at its middle (leading to a vanishing $P_{ion}$ contribution as defined in Eq.~\ref{eq:Pion}).  In the following, we compute  different CP chains polarization on both sides of  the gap closing point $\ta=\tb$ to check whether a topological phase transition occurs or not. 

As said above, due to the contribution of the lowest band, there is always a switch between the two Wannier center positions across the gap closing condition, and therefore a potential polarization jump, whatever the number of occupied bands, $M=1...,n-1$. According to Eq.~\ref{eq:Pq}, whether the Wannier center switch leads to a phase transition is triggered by a simple arithmetical condition : whether $0$ and $1/2$ belong to the same congruence class modulo $P_q$ (i.e whether 0 equals 1/2 modulo $P_q$). This question will be analyzed now, first through examples, and then in a general setting.

\section{Some examples }
\label{examples}

\subsection{Odd $n$ cases, with $\alpha=1/2$ and $\alpha=2/3$}
\label{sec:oddn}

The simplest nontrivial example beyond the alternating $AB$ chain is the $ABB$
chain, corresponding to $\alpha=1/2$ and $n=3$. It provides the minimal odd unit-cell case.
We also consider the $\alpha=2/3$ case, with $n=5$, corresponding to the chain $ABABB$.

\begin{figure}[t]
\centering
 \includegraphics[width=\columnwidth]{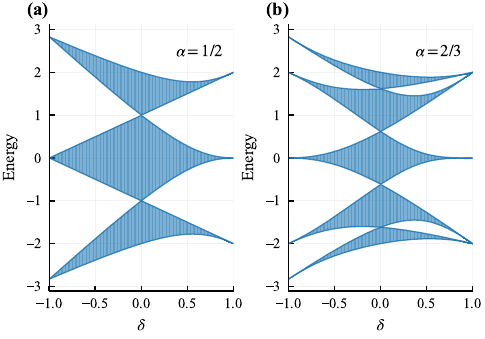}
\caption{
Band structure of two odd  $n$ cases :  $\alpha=1/2$, ABB (a) and $\alpha=2/3$, ABABB (b), plotted as a function of  $\delta$, with $\tb=1-\delta$ and $\ta=1+\delta$. At $\delta=0$ the gaps close.
}
\label{fig:oddn_spectrum}
\end{figure}

The canonical Bloch Hamiltonians are readily constructed from Eqs~[\ref{eq:Tj},\ref{eq:Hgeneral}], and  Figure \ref{fig:oddn_spectrum} displays the computed  spectra, with generically $n$ bands, as a function of $\delta=(\ta-\tb)/2$. As expected, the gaps close at $\ta=\tb$, separating gapped regions on both sides


As $\tb$ vanishes ($\delta=1$),  the $n$ bands turn into 3 (degenerate) energy values $E=\pm \ta=\pm 1$ and $E=0$. This is easily understood from the fact that the present CP chains, with $\alpha<1$ only shows isolated A bonds separated by sets of B bonds. At $\tb=0$ , the chain therefore decouples into isolated A bonds (leading to their bonding and anti-bonding eigenvalues $\pm \ta$), and isolated $B$ sites with vanishing energy. On the opposite side $\delta=-1$, the bands also converge toward a discrete set of eigenvalues. Since the chains do not have repeated $A$ bonds, there are no isolated sites in this limit, but rather a set of small "molecules" with variable number of sites separated by $B$ bonds. In the $n=3$ $ABB$ case, we just have isolated trimers (3 sites separated by two $B$ bonds). In the $n=5$ $ABABB$ case, there are isolated dimers and trimers, with their discrete set of eigenvalues.

One may ask how the situation evolves away from the isolated bonds limits, where Wannier states enter the description. Consider for instance a case with several (separated) $A$ bonds. At the $\tb=0$ limit we have uncorrelated states centered  at the $A$ bond mid-edges. As soon as $\tb>0$, these states interact through the $B$ bond interactions, and we expect a unique Wannier function centered at one of the two inversion centers. In the present odd $n$ case, the latter is located either at a site between two $B$ bonds, or at the mid-edge of $A$ or $B$ type. This information, if needed, cannot be extracted directly from Fig.~\ref{fig:oddn_spectrum}, and requires a finer analysis. What is clear however is that, away from the disconnected bonds limits, if the Wannier functions indeed change with the coupling strength, the inversion symmetry protects the Wannier center location, until the gap closes.

\begin{figure}[t]
\centering
 \includegraphics[width=\columnwidth]{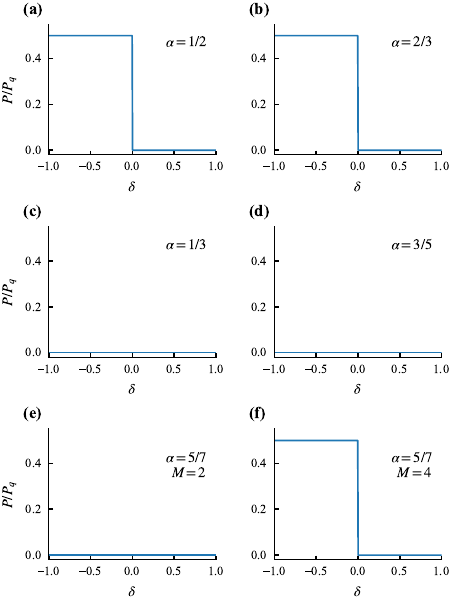}
\caption{
Polarization (divided by $P_q$) for several $(n,M)$ values. (a) $(n,M)=(3,1)$ for $\alpha=1/2$ ; (b) $(n,M)=(5,1)$ for $\alpha=2/3$ ;
(c)  $(n,M)=(4,1)$ for  $\alpha=1/3$; (d) $(n,M)=(8,1)$ for $\alpha=3/5$ (e) $(12,2)$ for $\alpha=5/7$ and (f) $(12,4)$ for $\alpha=5/7$ chains as a function of $\delta$. The odd $n$ cases shows a discrete jump of the polarization at the critical point, between $P=0$ and $P=P_q/2$, which is not always seen for the even cases.
}
\label{fig:polarization}
\end{figure}

We next compute the polarization. Since $n$ is prime for both the $\alpha=1/2$ and $\alpha=2/3$ cases, the quantum of polarization reads $P_q=1/n$ for any value of $M$. Figure~\ref{fig:polarization}-(a,b) displays the computed polarization (in units of $P_q$) for the two $n$-odd chains. They both display a jump at the gap closing transition $\ta=\tb$, indicating a (topological) phase transition between two inequivalent symmetry protected phases.

 At first sight, since polarization is related to Wannier center locations, we could be tempted to relate the polarization jump to the Wannier center jump from a site to a mid-edge. Some care has nevertheless to be taken with this statement as will be shortly shown in the even $n$ cases. A general analysis of this phase transition is postponed to Sec.~\ref{sec:general odd}.


\subsection{Even $n$ cases, with $\alpha=1/3$ , $\alpha=3/5$ and $\alpha=5/7$}
\label{sec:evenn}
The simplest even $n$ CP chain, beyond the alternating SSH $AB$ case, is the $ABBB$
chain, corresponding to $\alpha=1/3$ and $n=4$. 
We also consider here the $\alpha=3/5$ case, with $n=8$, and the $\alpha=5/7$ case, with $n=12$.

\begin{figure}[t]
\centering
 \includegraphics[width=\columnwidth]{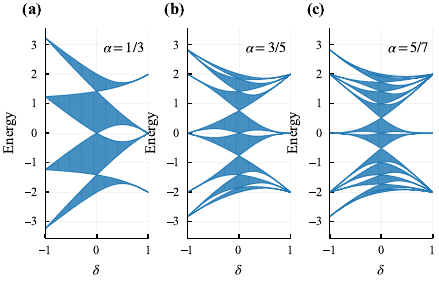}
\caption{
Band structure of two even $n$ cases : (a) $\alpha=1/3$ case, $ABBB$ chain; (b) $\alpha=3/5$ case (c)  $\alpha=5/7$ case,  plotted against $\delta$ (defined in the caption of Fig.~\ref{fig:oddn_spectrum}). 
 At $\ta=\tb$ the gaps close.}
\label{fig:evenn_spectrum}
\end{figure}

Figure \ref{fig:evenn_spectrum} displays the computed  spectra, with discrete eigenvalues in the limiting cases $\ta=0$ and $\tb=0$. The discrete eigenvalues are easily connected to the disconnected chain elements whenever one of the hopping term vanishes.


Here, when $\tb=0$, one again gets the same three eigenvalues found for the odd $n$ case, corresponding to $A$ dimers and isolated $B$ sites. Whenever $\ta=0$ one gets $B$ tetramers for $\alpha=1/3$, collections of $B$ dimers and trimers for the other two cases. Away from these extremal points, bands and gaps are formed, the latter persisting until the critical point is reached (here at $\delta=0$).

So at this point, we find no qualitative difference between the odd and even cases, as far as the energy spectrum is concerned. There is however a  difference with respect to the Wannier center locations on both sides of the gap closing point: here, as discussed in Sec.~\ref{sec:inversion}, they are both pinned at mid-edge centers, one at an $A$ bond center and the other, distant by one half unit cell, at a $B$ bond center. 

Fig.~\ref{fig:polarization}-(c,d,e,f) shows the polarization (in unit of $P_q$ for the Fermi level in the first gap (c,d) and for $\alpha=5/7$ in the second (e) and fourth gap (f). We find some differences : No jumps are found at the transition point in (c,d,e), indicating an absence of phase transition, while the case $(n,M)=(12,4)$ shows a polarization jump, indicating a phase transition.

Notice that, here also, there is a jump of the Wannier center, shifting from a $A$-bond center to a $B$-bond center across the $\delta=0$ point. This proves already  that the presence of a Wannier center discrete shift is not a sufficient condition for a polarization jump. This point will be clarified in Sec.~\ref{sec:general even}.

\section{General classification}
\label{sec:general}

We now discuss the general case of a rational slope $\alpha=p/q$ and a unit
cell containing $n=p+q$ sites, in search for the occurrence of a topological phase transition.

The analysis of the preceding sections suggests that such a transition depends not only on the parity of $n$, but also on $M$, the number of occupied bands. The crucial point regards the definition of $P_q$, and the condition under which the polarization lattices on both sides of the critical point coincide or not.

This can be rephrased as whether the two calculated possible values $0$ and $1/2$ are congruent modulo $P_q$, which translates into a simple  arithmetical criterion for $(n,M)$:

\begin{itemize}
 \item 
 If $0=1/2 \pmod{P_q}$, there is no phase transition across the gap closing point.
 Two cases : 
 
 (i) $n$ is even and $M$ odd,
 
 (ii) $n$ is even , $M$ even and $\gcd{(M,n)}$ divides $n/2$.

\item
If  $0\neq 1/2 \pmod{P_q}$, there is a topological phase transition.
 Two cases : 
 
 (i) $n$ is odd for $\forall M \in \{1,n-1\}$,
 
 (ii) $n$ is even , $M$ even and $\gcd{(M,n)}$ does not divide $n/2$.

\end{itemize}
An equivalent (more compact) formulation reads  : A topological transition occurs if and only if $n/\gcd(M,n)$ is odd. When $n/\gcd(M,n)$ is even, the two insulating regimes are equivalent and no topological transition occurs.
Let us now discuss this general result separately for $n$ odd and even.

\subsection{The odd $n$ case} 
\label{sec:general odd}
For general odd $n$, the two insulating regimes involve Wannier centers located on different types of inversion centers, typically site-centered versus bond-centered. 

On one side of the critical point, the Wannier center of the lowest band is located at the (inversion symmetric) site separated by two $B$ bonds. On the other side it will sit at the center of a $A$ or a $B$ bond.  For intermediate bands, Wannier centers do not shift at the critical point, as discussed in Sec.~\ref{sec:indicator}.

From a physical point of view, it is natural to expect that two phases such that their Wannier centers sit at a site in one case and at a bond center in the other case cannot be topologically equivalent. In terms  of polarization lattice coincidence, this is also quite evident. For $n$ odd, the quantum of polarization $P_q$ as given by Eq.~\ref{eq:Pq} is the inverse of an odd number, and $0$ and $1/2$ are not congruent modulo an inverse of an odd number.

The conclusion is therefore that, for odd $n$, the two phases have different polarization $P=0$ and $P=P_q/2$ and that there is a TPT at $t_A=t_B$.

\subsection{The even $n$ case}
\label{sec:general even}
This case will prove more subtle to analyze. For even $n$, even though the two insulating regimes  show a shift of Wannier centers located respectively on $A$ and $B$ bond centers, this does not necessarily lead to a phase transition across the closing gap point.

Notice that this Wannier center shift is already observed for the SSH $AB$ case, and it was shown  that polarization does not change, both phases being equivalent up to a translation by one half of the unit cell length \cite{FuchsPiechon2021}. For generic even-$n$ chain, we will have a similar conclusion for $M$ odd, although we will need to invoke a more complicated argument.

For the even $n$ case, we find that in most  cases, there is no TPT and the polarization vanishes. While in few cases, there is a TPT between two phases with $P=0$ and $P=P_q/2$. Asymptotically when $n$ is even and $n\to \infty$, the probability of no TPT tends to 1.

\subsubsection{$C$-path approach}
\label{sec:C-path}

The general idea is to find a continuous path between $t_B < t_A$ and $t_B > t_A$ without going
through the gap closing point $t_B = t_A$, while maintaining the chain inversion symmetry (to keep the Wannier centers at symmetry points along the path). This is done by transforming the $(A,B)$ chain into a ternary $(A,B,C)$ one, with the $C$ bond hopping parameter $t_C$ continuously changing from $\tb$ to $\ta$.  The detailed construction is presented in Appendix~\ref{app: C path}. 

Here we limit the presentation to a simple example. We choose $p=1$,
$q=3$ so that $n=4$ and the unit cell is $ABBB$. The continuous path (which we call the “$C$ path”) proceeds in 5 steps illustrated on a single unit cell (inversion centers are underlined):
\begin{itemize}
\item
1) $\underline{A}B\underline{B}B$ with $t_B < t_A$.
\item
2) $\underline{A}C\underline{B}C$ with $t_B < t_C < t_A$ is a ternary chain that preserves inversion symmetry.
The $C$’s (in step 2) are initially $B$’s (in step 1) and progressively turn into $A$’s (in step 3).
\item
3)  $\underline{A}A\underline{B}A$ with $t_B < t_C = t_A$.
\item
4)  $\underline{B}B\underline{A}B$ with $t_A < t_B$ after relabeling $A$ in $B$ and $B$ in $A$.
\item
5)  $\underline{A}B\underline{B}B$ with $t_A < t_B$ after spatially shifting the chain by half a unit cell. In the end, 5) is identical to 1) except that $t_B < t_A$ has turned into $t_A < t_B$ without ever
crossing $t_B = t_A$.
\end{itemize}
What remains to be checked is whether electronic gaps remain open or not in this process. In the
following, we separately analysis gaps in cases where $M$ is odd or even.

\subsubsection{The odd $M$ case}
\label{sec:general n even M odd}

This non-closing condition is discussed in Appendix~\ref{app:gap_analysis}. It is shown that, counting gaps from the lower one, for Fermi levels in gaps labeled by $M$ odd (with gap openings at $k=\pi$), the non-closing condition is simply achieved. Basically, the argument is that the perturbation introduced by $C$’s has a periodicity of half a unit cell and therefore Fourier components $q$ that are multiple of $4\pi$. In degenerate perturbation theory, it only affects wavevectors $k$’s that are multiple of $2\pi$ (as $q=2k$). Therefore, it can not affect odd gaps that correspond to $k=\pi$, but can affect even gaps that correspond to $k=0$.
Our argument cannot be applied to the $M$ even cases, for which the gaps open at $k=0$.

Back to the above arithmetical criterion, it is clear that if $M$ is odd, $\gcd(M,n)$ is odd; since it divides the even integer $n$, it also divides $n/2$, leading to an absence of TPT. Our ``$C$ path" construction provides here an independent proof for even $n$/odd $M$.

\subsubsection{The even $M$ case}
\label{sec:general n even M even}

For even $M$ we numerically find that gaps close along the continuous $C-$path, and we can therefore only rely on the above arithmetical criterion related to a peculiar polarization quantum $P_q$. Notice that examples given in Sec.~\ref{sec:evenn} agree with this criterion. Despite the fact that the gap closes along the $C$-path, in most cases, the polarization does not change between $t_A<t_B$ and $t_A>t_B$ so that there is no TPT. We shall see below that these $P_q$ values shows up when computing the polarization of open boundary CP chains.

\subsection{Inversion-symmetric chains beyond CP}

Let us make a general comment on the occurrence of a topological phase transition in 1D inversion symmetric chains. In this paper, we focused on an off-diagonal tight-binding model applied to CP chains, which by construction possess inversion symmetry. However, our arithmetical criterion does not involve the pair ($p,q$), which contains the subtlety of the chain unit cell sequence, but only $n$, the number of sites in the unit cell  and $M$, the discrete band filling. As a consequence,  this result holds for more generic inversion-symmetric chains.

 A first such example is described in Appendix~\ref{app:diagonal term}, where we have added a diagonal term in the Hamiltonian to differentiate sites surrounded by an $A$ and a $B$ bond with those surrounded by two $B$ bonds. This choice preserves inversion symmetry, but the  gap closing condition now depends upon $M$. As expected, we find that our TPT classification still applies.

But this classification applies for generic inversion symmetric chain (with both diagonal/off-diagonal interactions). The only difference in the analysis is the possible occurrence of adjacent  $AA$ bonds. This may allow for cases where the Wannier center jumps occur between two different site positions, a case that does not happen for CP chains.

\section{Polarization of open CP chains}
\label{sec:edges}

The bulk-edge correspondence predicts protected in-gap boundary states  for symmetry-protected topological (SPT) phases whose protecting symmetry is preserved at the boundary \cite{Fisher2013}: Time-reversal symmetry provides an example. Consider an interface between a topologically non-trivial material occupying the half-space ($x<0$) and a topologically trivial medium for ($x>0$). Since the corresponding bulk topological invariants differ on the two sides of the interface, they cannot be continuously connected while preserving both the bulk energy gap and the protecting symmetry. 
Consequently, if the protecting symmetry is maintained, the bulk gap must close locally at the interface.
 This local gap closing manifests itself as boundary states with energies lying inside the bulk gap.

The situation is different for one-dimensional inversion-protected phases, which are the subject of the present work. A physical boundary necessarily breaks inversion symmetry, so that the symmetry protecting the bulk invariant is no longer present at the edge. As a consequence, the bulk-edge correspondence does not predict symmetry-protected in-gap edge states. Any boundary states that may appear therefore depend on the microscopic termination of the chain rather than on the bulk topological invariant alone. A peculiar case concerns edge states in CP chains with rescaled slope $1/n$ which are described in appendix~\ref{app:ABn}.

Here we shall mainly consider open chain properties which can be put in correspondence with bulk TPTs.  The relevant form of the bulk-edge correspondence is usually expressed through the surface charge theorem, which relates
the boundary charge $Q$ to the bulk polarization $P_{PBC}$ modulo an appropriate quantum of polarization~: $ Q=P_{PBC} \pmod{P_q}$ \cite{KingSmith1993}.  For  large OBC chains of length $L$, and if the electric charge localization length is much smaller than $L$, we expect that the open boundary polarization per unit length, $P_{OBC}$,  equals the total electric dipole $Q \times L$ divided by $L$, and therefore $P_{OBC}=Q$. Therefore the surface charge theorem can also be written as $P_{OBC} = P_{PBC}$ mod $P_q$.

Here we shall compare $P_{OBC}$ for chains differing by a shift in the unit cell. More precisely, we cut a large PBC chain with $L$ repeated unit cells at $n$ inequivalent positions in one unit cell. We get $n$ related OBC chains with $L \times n$ sites.

Notice that, in contrast with the PBC case, the OBC polarization calculation is now direct, without invoking Resta position operators, Zak phases and symmetry indicators. In addition, in an open chain, translation symmetry is broken and there is therefore no polarization quantum. We just get the spectral properties from a numerical diagonalization, and add the contribution of each occupied state via the occupation probability of each site $i$ in the state $\Psi_j$ : $- x_i |\Psi_j(i)|^2$ (recall that we have fixed the electron charge to $-1$). The ionic part vanishes, as a consequence of placing the coordinate origin at the OBC chain center.

\begin{figure}[htbp]
\centering
 \includegraphics[width=\columnwidth]{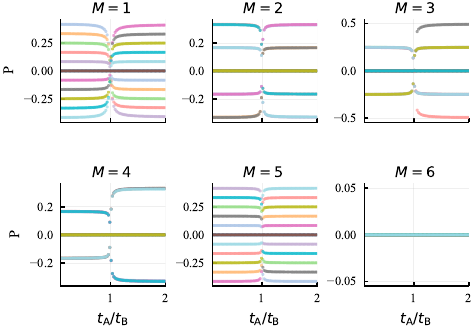}
\caption{
Polarization for OBC chains made of $L=100$ repeated unit cells with respect to the  number $M$ of occupied bands, for the different inequivalent shifts in the unit cell. Here the case $\alpha=5/7$ is shown. We observe that, for a given $M$, the different $P$ curves indeed differ by $P_q = \gcd(M,n)/n =  1/12, 1/6, 1/4, 1/3, 1/12 , 1/2$ for $ M= 1, 2, 3, 4, 5, 6$.}
\label{fig:OBC_polarization}
\end{figure}

Fig~\ref{fig:OBC_polarization} shows polarization calculated for an OBC chain with $n=12$, variable $M$, and with $12$ different configurations for the shifted unit cells. It is found that the $n$ different $P_{OBC}$ differ by a multiple of the polarization quantum $P_q$, although the latter is a quantity associated to shift of the unit cell definition in the PBC context and does not enter the OBC calculation. This is a manifestation of a bulk/boundary correspondence. Note however that, for one choice of unit cell, the open chain has global inversion symmetry. As a consequence, $P=0$ for all $t_A/t_B$. This particular choice should be discarded in order to obtain $P_q$. For example, when $M=4$, $P_q=1/3$ and there is a TPT between $P=P_q/2$ when $t_A/t_B<1$ and $P=0$ when $t_A/t_B>1$.
In summary, the OBC calculation reveals the correctness of the $P_q$ obtained in the PBC case.

\section{Conclusions and perspectives}
\label{sec:conclusion}

We have introduced a broad family of inversion-symmetric binary tight-binding chains generated by CP and described the main features expected in their spectral and topological properties. The construction provides a natural setting in which to study the interplay between inversion symmetry, Wannier centers, electric polarization and boundary states.

Across the gap closing $\ta=\tb$ point, the total Zak phase always shows a discrete jump, but this is not a sufficient condition for having a topological phase transition. Our main result here is that the occurrence of such a transition depends on an arithmetical criterion involving  the number of sites in the unit cell $n$ and the number of filled bands $M$. A surprising result is that it does not depend on the subtle sequence $S(\alpha)$ defined by the pair $(p,q)$, as long as inversion symmetry is preserved. Our result is therefore more general and applies to any periodic binary chain with inversion symmetry. Notice that the simple non-CP chains with unit cells $AABB$ and $AAABBB$ already display different topological properties for even $M$. The focus on CP chains is a way to ensure this symmetry. It also allows for an independent proof of the absence of phase transition for $n$ even and $M$ odd (see Sec.~\ref{sec:general even}). 

Let us  mention here that an interesting property also occurs near the gap closing point for the set of off-diagonal CP chains considered in the present work. In the SSH case, the $\theta=1/2$ chain is a simple model for Peierls dimerization, which occurs at half filling in the periodic standard chain. This property has been extended  by C. Sire~\cite{Sire1991} : For any band filling (here rational, to stay in the context of the present work), there exists a (eventually small) modulation of the periodic chain, modeled on a $S(\alpha)$ sequence, which has lower energy compared to the periodic chain.

 We have already checked, in appendix \ref{app:diagonal term}, the generality of our result on TPT occurrence by including inversion-preserving diagonal terms. For $n>2$, such terms can be introduced in a natural way by assigning on-site energies $\pm \lambda$ according to the local bond environment,  distinguishing sites surrounded by $A$ and $B$ bonds from those surrounded by two $B$ bonds. In this context, the case $\ta=\tb$ displays a remarkable feature, already noted in ref\cite{SireMosseri1990}, that for any $\lambda$ value, the spectrum displays one gap closing, providing a new potential topological phase transition. Upon computing the polarization, we show that, as expected, our arithmetical TPT criterion applies exactly in that case.

A first perspective should be to study what happens at the vicinity of irrational slopes, and how to characterize the transitions occurring at $\ta=\tb$. The effect of inversion symmetry in the Fibonacci chain has been addressed in Ref.~\cite{levy2016}, who analyzed the existence  of palindromic occurrences with respect to the phason coordinate in $E_\perp$. In the present context, where the focus is on the existence of TPT at the gap closing point, one way would be to study a sequence of best approximants in the Farey sequence, as for the quasiperiodic Fibonacci chain and its rational approximants based on successive ratio in the Fibonacci sequence. Approaching the irrational slope, a complex pattern of bands subdivide step by step, and eventually leads to a singular continuous spectrum~\cite{Bellissard1992} (see also Ref.~\cite{Beckus2026} for a more recent work on CP chains with diagonal modulations). But, according to our present arithmetical criterion for the TPT occurrence, the path to the irrational slope may lead to some additional complexity, depending on the Fermi level location: Simple sets of rational approximants sequences (even taken from the best approximants) may display, or not, a TPT while navigating along the sequence. A precise (and unique) characterization of the transitions occurring at irrational slopes should take this additional complexity into account which also amounts to revisiting the concept of quantum of polarization at the quasiperiodic limit.
A recent study~\cite{Moustaj2025} has considered the problem of the electronic polarization of aperiodic chains with inversion symmetry, but did not take ions into account and also not the polarization quantum of periodic approximants.

Another possible direction is the study of higher-dimensional cut-and-projection structures, such as codimension-one two-dimensional systems and their periodic approximants~\cite{VidalMosseri2000}. Such systems also possess inversion symmetry, although the conditions under which  gaps appear and/or close remain to be analyzed.

\acknowledgements
This work is supported by the National Natural Science Foundation of China (Grant Nos. NSFC-12504192 and 12374046) and the Shanghai Science and Technology Innovation Action Plan (Grant No. 24LZ1400800). J.-N. F. thanks Fred Piéchon and Malik Ayachi for previous collaboration on the polarization of inversion-symmetric chains, and R. M. thanks Clément Sire for previous collaboration on the spectral properties of CP chains.

\appendix

\section {Analysis of the CP butterfly skeleton}
\label{app:farey discussion}

\begin{figure}[htbp]
    \centering
    \includegraphics[width=\linewidth]{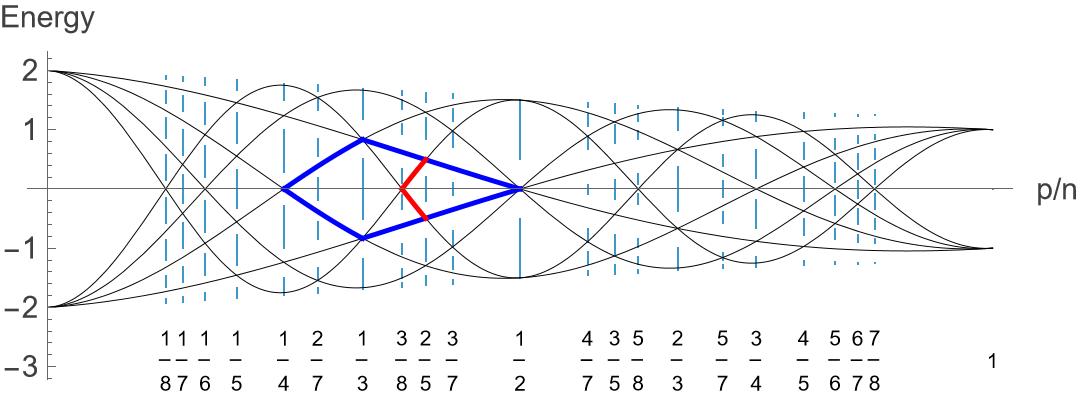}
     \caption{Skeleton of the CP butterfly built from a set of scaled cosine curves (see Eq.~\ref{eq: cosine curves})  threading the gaps. Here, the rational values are taken in the Farey series $F(8)$ (whose values appear in the lower part of the figure), and 4 cosine curves are needed. Also drawn is the kite (with four blue sides) built on the triple $(1/4,1/3,1/2)$, which contains a smaller kite (limited by two red and two blue sides) built on the triple $(3/8,2/5,1/2)$. }
    \label{fig:skeleton}
\end{figure}

To analyze the CP butterfly, we can go to the perturbative limit where $\ta \approx \tb$ as was done in Ref.~\cite{SireMosseri1990}, who gave a general formula for the gap openings and widths for arbitrary value of $\theta=p/n$.

The gaps are ordered (roughly, with small corrections) according to decreasing size with  an integer $j$, and reads, for a given  $\theta$ value :

\begin{equation}
E_G(j,\theta)= \pm 2\ \bar{t}\ \cos \pi j \theta,
\label{eq: cosine curves}
\end{equation}

 where $\bar{t}=  \tb(1-\theta) +\ta\theta $ is the average hopping term and  $j$ labels the gaps, their size roughly decreasing with increasing $j$. 

In fact, these cosine curves thread the whole pattern and form the skeleton of the full CP butterfly. If one plots the band pattern for all rational in a Farey series $F(m)$, the $m$ cosine curves given in Eq.~\ref{eq: cosine curves}, for $j=1 \cdots m/2$ and both signs counted separately (we take $m$ even for simplicity), only cross in the gaps or at the band edges for these rational values (see Fig.~\ref{fig:skeleton}).

A prominent feature of the CP butterfly is the emergence of self-similar spectral (rounded) quadrangular shapes, which we call ``kites", defined by a triple of rational numbers whose middle one bounds the largest band, and limited  by (roughly)  cosine curve pieces. 

The self-similar hierarchy of kites reflects a corresponding hierarchy in the underlying $S(\alpha)$  sequences. Rational structures (with larger $n$) close to a given $\theta$ value may be viewed as long domains reproducing the parent sequence, separated by sparse periodically spaced discommensuration defects whose density decreases with the distance to $\theta$. The recursive organization of these defects provides a geometrical interpretation of the butterfly spectral self-similarity.

Let us describe as an example what happens on the left of the central large gap (at vanishing energy)  for $\theta=1/2$ (see Fig.~\ref{fig:farey}  for the details and \ref{fig:skeleton} for a schematic view), with the triple $(1/4,1/3,1/2)$,  corresponding to the alternating $AB$ SSH model. We see an emerging kite pattern starting as a sharp mid-gap level (which exists as soon as $\theta<1/2$), which further widens. Since states near zero energy dampen rapidly in the $AB$ gapped regions, we conclude that these widening bands originate from this periodic array of defects, here a periodic insertion of $B$ bonds, leading locally to $ABB$ patterns, which become closer as $\theta$ departs from $1/2$, until these patterns become eventually first neighbours, building the $\theta=1/3$ ($ABB$ unit cells) normal bands. In the range $1/3<\theta<1/2$ all sorts of unit cells  mixing $AB$ and $ABB$ patterns are found with their respective normal bands. The same analysis can be carried out for the triple $(3/8,2/5,1/2)$, defining the  embedded kite in  (Fig.~\ref{fig:skeleton})  and more generally for any other kite in the butterfly.

\section{Analysis of the even $n$ case}
\label{app:even n case}

\begin{figure}[t]
\centering


\begin{minipage}{0.48\columnwidth}
\centering
\includegraphics[width=\linewidth]{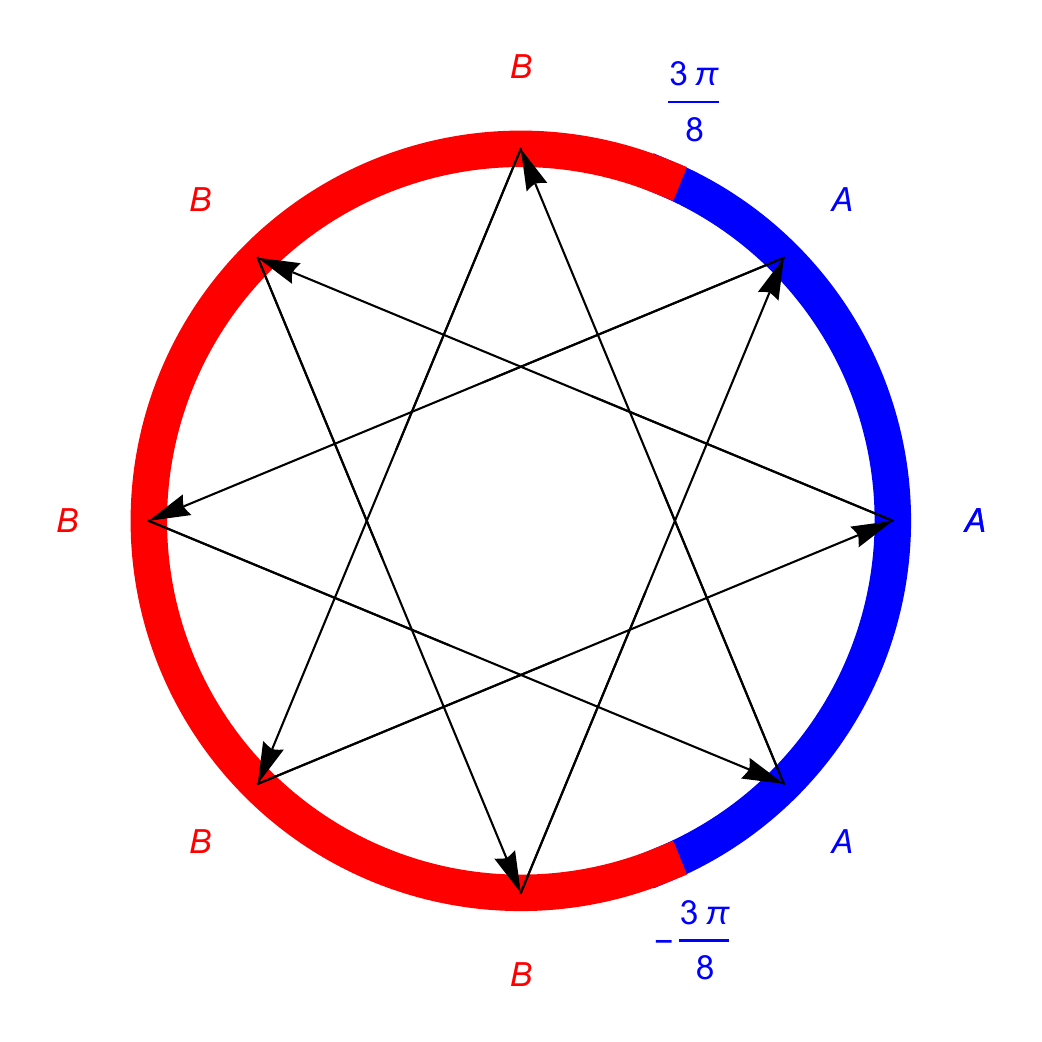}

\vspace{1mm}
(a)
\end{minipage}
\hfill
\begin{minipage}{0.48\columnwidth}
\centering
\includegraphics[width=\linewidth]{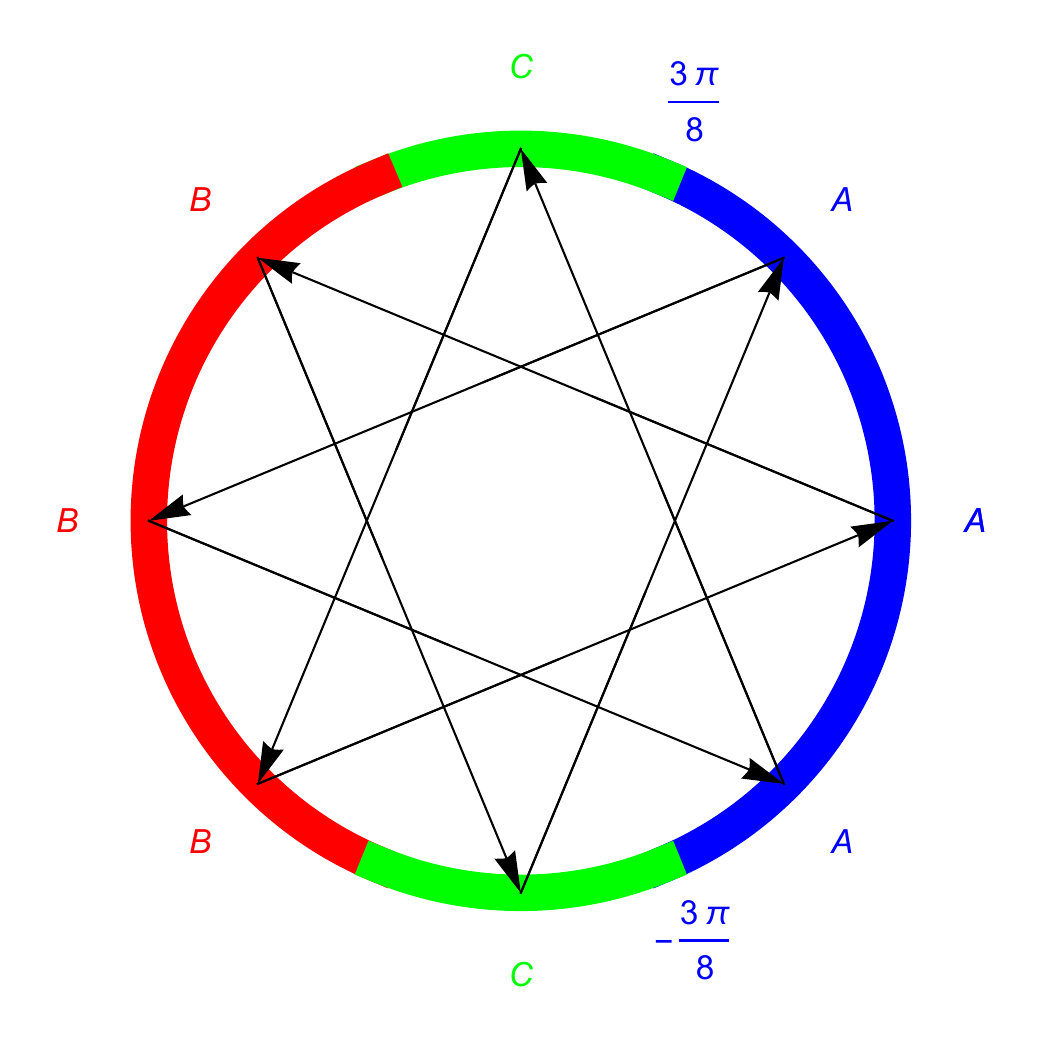}

\vspace{1mm}
(b)
\end{minipage}

\vspace{3mm}


\begin{minipage}{0.48\columnwidth}
\centering
\includegraphics[width=\linewidth]{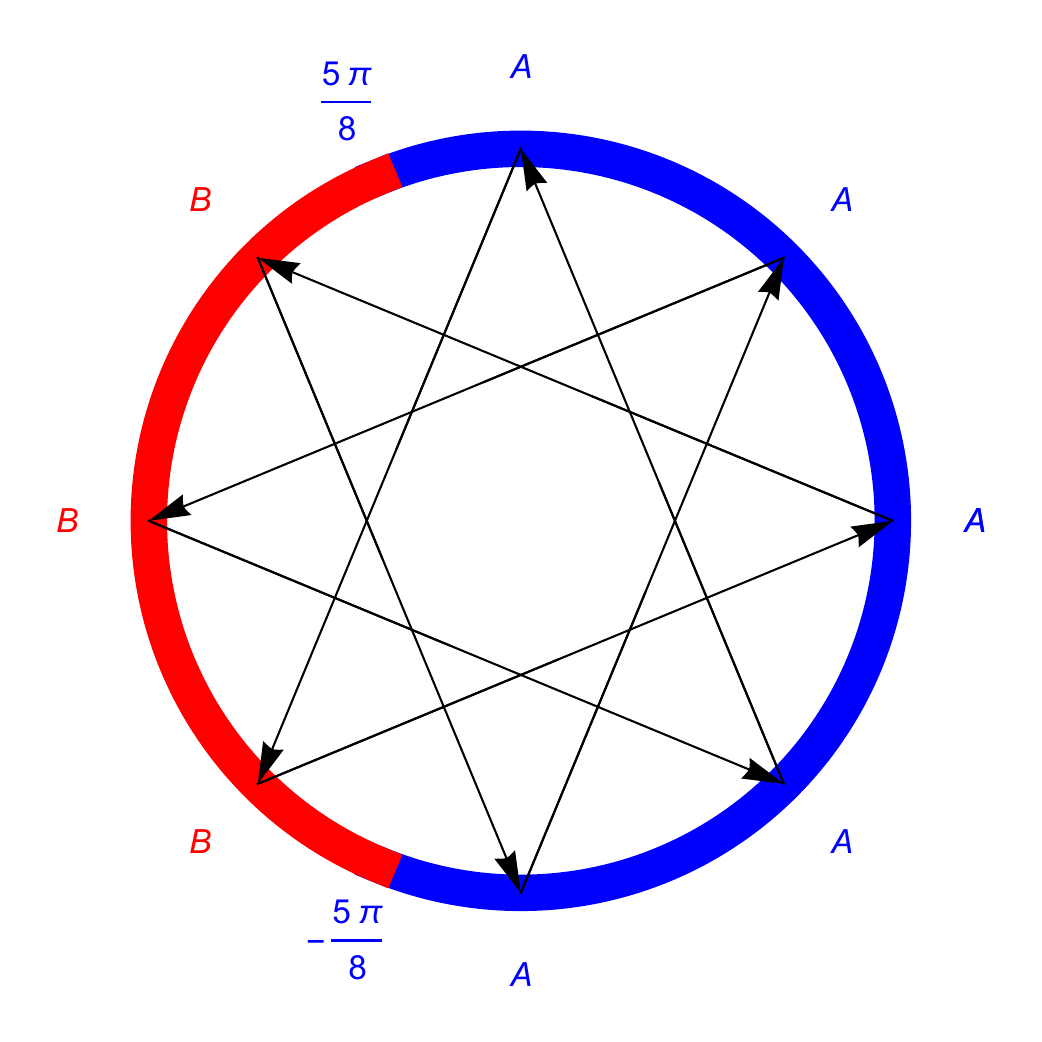}

\vspace{1mm}
(c)
\end{minipage}
\hfill
\begin{minipage}{0.48\columnwidth}
\centering
\includegraphics[width=\linewidth]{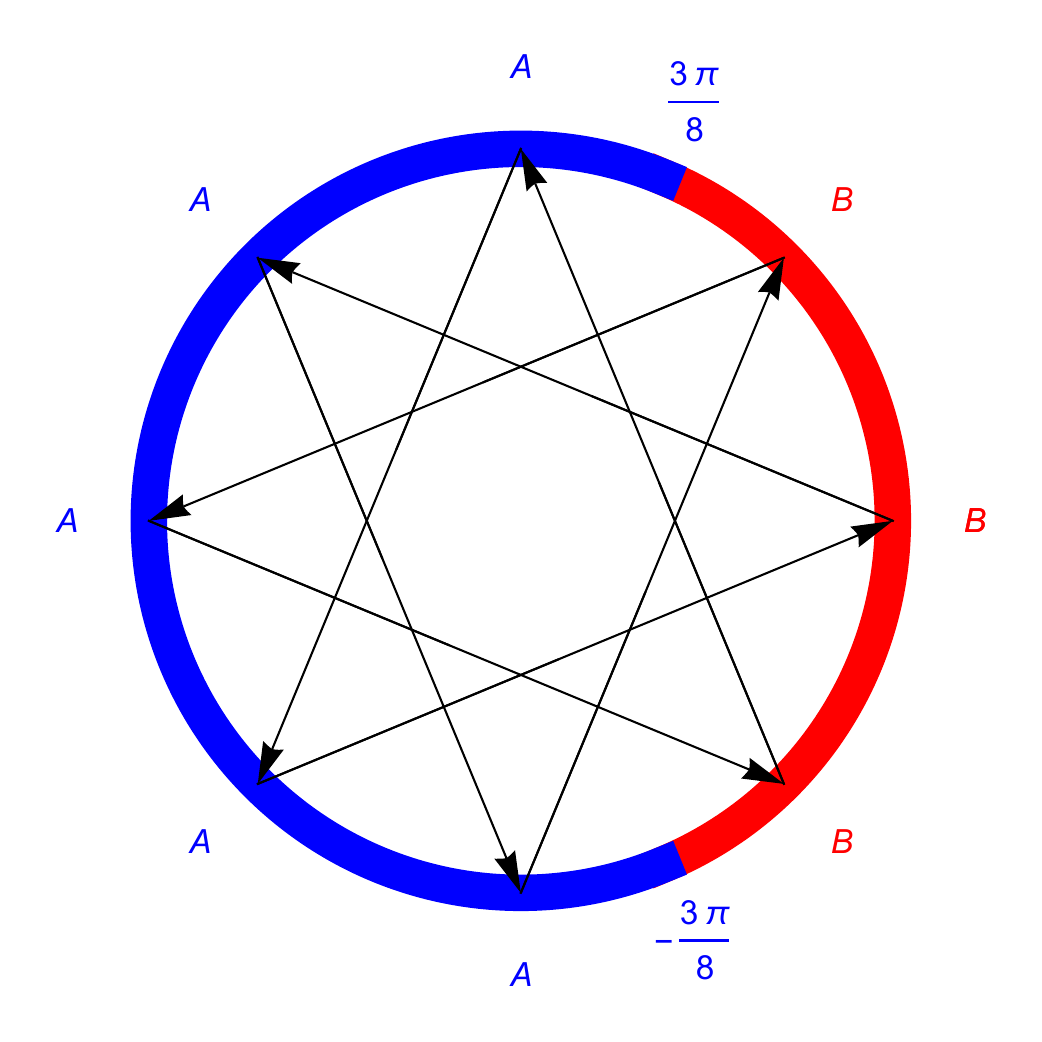}
\vspace{1mm}
(d)
\end{minipage}

\caption{
(Color online)
Illustration of the passage from a $S(\alpha)$ sequence to a $S(1/\alpha)$, in the case $(p,q)=(3,5$): (a) standard circle rotation algorithm, with a stellated octagon. The sequence $ABBABABB$ is generated; (b): after introducing the window $W_C$, two $B$ bonds are transformed into 2 $C$ bonds, leading to the sequence $ABCABACB$;(c) changing further $C$ into $A$, leading to the sequence $ABAABAAB$; (d) upon rotating the windows by $\pi$, corresponding to a translation of half a unit cell distance in the periodic chain, one gets the sequence $BAABABAA$, which is identical to the initial one in (a), upon relabelling $A\leftrightarrow B$.}
\label{fig:circle rotation}

\end{figure}

\subsection{``$C$ path" construction.}
\label{app: C path}

We now detail the construction of a particular continuous path in the unit cell bond hopping parameters, which starts from a $S(\alpha)$ Hamiltonian and ends with a $S(1/\alpha)$ Hamiltonian. As discussed above, this amounts to exchange $A$ and $B$ bonds in the unit cell. But we want this path to go from $\tb>\ta$ to $\ta>\tb$ without reaching the gap closing critical point $\ta=\tb$, and we cannot therefore use a simple continuous global exchange $\ta \leftrightarrow \tb$.

 This is done here by  selecting a set of $B$ bonds into new bonds, called $C$ bonds, whose hopping term is continuously changed from $\tb$ to $\ta$, and  are such that inversion symmetry is preserved along the path. 
 
 We use here a ``circle rotation" algorithm for the bond sequence, tuned to be equivalent to CP site selection. It is illustrated in Fig.~\ref{fig:circle rotation} in the $\alpha=3/5$ case. For generic (co-prime) $\{p,q\}$ a unit radius circle is divided into two semi-open angular intervals : $W_A=[-\pi p/n,\pi p/n[$ and $W_B=[\pi p/n, \pi (2- p/n)[$; The bond sequence $S(\alpha)$ is then generated by sequential rotation of angle $\beta=2\pi p/n$, starting for instance at the point $x(1)=(1,0)$; The repeated points $x(j)=( \cos (j-1) \beta,\sin (j-1)\beta)$ form a stellated n-gon. Since $n$ is even, the point pattern has inversion symmetry. Now, for each point in the sequence, a letter $A$ or $B$ is chosen, according to whether $x(j)$ fall in interval $W_A$ or $W_B$ (see Fig.~\ref{fig:circle rotation}-left). The obtained sequence is identical (up to circular permutation) to the CP $S(\alpha)$ sequences considered in this paper.
 
 We now generate a new ternary sequence with bonds $(A,B,C)$ by introducing a (double) interval $W_C$, which is the union of $[\pi p/n, \pi (1-p/n)[$ and $\pi (1+ p/n),\pi (2-p/n)[$. $W_C$ is the intersection of $W_B$ with $W_B$ rotated by $\pi$ (translation of one half of the unit cell in corresponding periodic set). In terms of the initial $(A,B)$ sequence, it contains all pairs of $B$s distant by $n/2$.
 
 The next step consists of continuously changing the $t_C$ Hamiltonian hopping terms from $t_B$ to $t_A$. At the end of this process, we can replace $C$ bonds by $A$ bonds, and recover an $(A,B)$ binary sequence. to characterize this sequence we can redefine the windows $W_{A/B}$ into new ones $W^\prime_{A/B}$ corresponding to the new selected letters (see Fig.~\ref{fig:circle rotation}-c). To get  $W^\prime_{B}$ we must subtract $W_C$ from $W_B$ :  $W^\prime_{B}= W_B \setminus W_C$. Accordingly, we get $W^\prime_{A}=W_A \cup W_C$.
 
 We now clearly see that $W^\prime_{B}$  is equal to $W_A$ rotated by $\pi$ and $W^\prime_{A}$ is   equal to $W_B$ rotated by $\pi$.  This proves that the final sequence corresponds to the initial one with an exchange $A \leftrightarrow B$, and a cyclic permutation of $n/2$ (associated with the above $\pi$ rotation of the respective intervals). We have therefore achieved our initial goal by modifying only a small set of $B$ bonds into $A$ bonds.
 
 A point which will prove important in the next section is to remark that the set of $C$ sites correspond to diametrally opposed points on the circle, and therefore to pairs of points separated by half a unit cell in the repeated periodic chain.

\subsection{ Gap non-closing condition}
\label{app:gap_analysis}

\subsubsection{gap non-closing condition along the C path}
\label{app:gap non closing}

\begin{figure}[htbp]
    \centering
    \includegraphics[width=\linewidth]{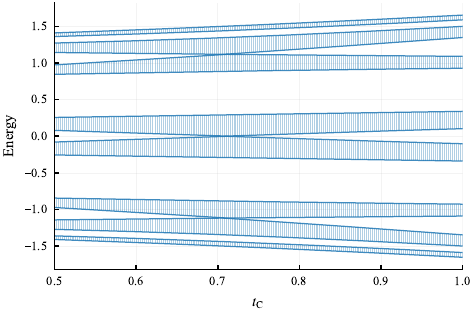}
    \caption{Spectrum support along the $C$ path, in the case $\alpha=3/5$, as $t_C$ runs from $t_B=0.5$ to $t_A=1$. Odd gaps never close, while even gaps close at $t_C = \sqrt{t_A t_B}$.}
    \label{fig:gap path}
   
\end{figure}

The above described C path continuously transforms a $S(\alpha)$ sequence to $S(1/\alpha)$, for a fixed pair ($t_A,t_B)$. As said in the main text, this implies that  the end sequence is equivalent to the initial one upon translation by half a unit cell and relabeling $A$ into $B$ and $B$ into $A$.

Our aim to construct this $C$ path was to better understand the numerical observation that for $n$ even and odd $M$, no topological phase transition was observed when crossing the critical point $\ta=\tb$. The idea is to show that for a path conserving the inversion symmetry, and avoiding the critical point both spectra are connected without the occurrence of a gap closing.

As an example, Fig.~\ref{fig:gap path} shows the spectrum support along this path, for the case $\alpha=3/5$. We observe that odd numbered gaps (1,3,5,7) never close, while eventually even numbered ones close.

This can be understood by the following remark. The Hamiltonian along the path $C$ has two terms : a constant part associated to the $A$ bonds and the $B$ bonds not transformed into $C$ bonds. We call this part the unperturbed Hamiltonian. In addition, we have the varying terms associated with the $C$ bonds, which we call the perturbation. By construction this perturbation is periodic, with period $1/2$, and therefore has Fourier components at integer multiples of $4 \pi$ (recall that we have fixed unit cell lengths to 1). It therefore affects the constant part only at $k=0$ and $k =\pm 2 \pi$, where gaps can close or open. So, this leaves the component $k=\pi$ unaffected. Starting at $x=0$ in a gaped phase at $k=\pi$, the perturbation cannot close this gap. For gaps $k=0$, the path $C$ shows a gap closing, which occurs at the value
$t_C= \sqrt{\ta\tb}$ (see Fig.~\ref{fig:gap path}).

\subsubsection{an alternative description}
\label{app:mirror_perturbation}

This $C$-path non-closing gap behaviour can also be described in the following way, again in the perturbative regime $\ta \sim \tb$, by focusing on the two quasi-degenerate states  at $k=\pi$ (symmetric or anti-symmetric with respect to an inversion center placed at the center of the unit cell).

We can write the Hamiltonian of the  chain as \(H = H_0 + \delta H\), where $H_0$ is the periodic chain Hamiltonian with uniform hopping amplitude $t=(\ta+\tb)/2$ and $\delta H$ the ``perturbation", with hopping terms $\delta t = \pm (\ta-\tb)/2$ whose signs follow  the particular sequence $S(\alpha)$.

For $\ta$ close to $\tb$, we perform a degenerate perturbation analysis at the Brillouin zone boundary \(k = \pi\) of the folded Brillouin zone (recall that we have set the unit cell length $a=1$ throughout this paper) and write it in the basis formed by the symmetric and anti-symmetric states  $(\ket{\psi_+},\ket{\psi_-})$, with respect to a unit cell inversion center.

Since \(\delta H\) respects the inversion symmetry \(I\) (i.e., \([\delta H, I]=0\)), and \(\ket{\psi_\pm}\) are eigenstates of \(I\) with distinct eigenvalues (\(+1\) and \(-1\)), the off-diagonal matrix elements vanish by the symmetry selection rule: \(\braket{\psi_+ | \delta H | \psi_-} = \braket{\psi_- | \delta H | \psi_+} = 0\). Therefore, in the \(\{\ket{\psi_+}, \ket{\psi_-}\}\) basis the $2\times 2$ perturbation matrix is diagonal and takes the form:
\begin{equation}
    \delta H = \epsilon(\alpha) \ket{\psi_+}\bra{\psi_+} - \epsilon(\alpha) \ket{\psi_-}\bra{\psi_-}.
    \end{equation}

The explicit expression of $\epsilon(\alpha)$ is not needed in the following.
Following the construction of the $C$ path introduced above, we now seek a path that avoids the gap-closing point $\ta=\tb$, corresponding here to $\delta H = 0$.

The key observation is that $\ket{\psi_+}$ and $\ket{\psi_-}$ are related by a translation of half a unit cell.
An easy calculation indeed leads to $\ket{\psi_+} = \ket{\psi_-^{1/2}}$, 
where \(\ket{\psi_-^{1/2}}\) is the wave function of \(\ket{\psi_-}\) shifted left by half a period. Similarly, we get that \(\ket{\psi_-} = \ket{\psi_+^{1/2}}\).

We also introduce the shifted form $\delta H_{1/2}$ for the perturbation. 

We fix $\ta \neq \tb$, to avoid the gap closing point, and introduce a continuous  path for the the Hamiltonian in the form \(H(x) = H_0 - \tau\delta H + (1-\tau)\delta H_{a/2}\), with $\tau$ running from $0$ to $1$,  and therefore \(H(0) = H_0 + \delta H_{1/2}\) and \(H(1) = H_0 - \delta H\). Since $H_0$ commutes with translations on the chain, we get that $H(0)$ is equivalent to $H_0 + \delta H$ up to a translation by $1/2$

Now, using the above relations between $(\ket{\psi_+},\ket{\psi_-})$ and $(\ket{\psi_+^{1/2}}, \ket{\psi_-^{1/2}})$, the $ 2\times 2$ Hamiltonian in the (${\ket{\psi_+}, \ket{\psi_-}}$) basis stays diagonal, and independent of $\tau$:
\begin{equation}
\begin{aligned}
    H(x) &= E \left( \ket{\psi_+}\bra{\psi_+} + \ket{\psi_-}\bra{\psi_-} \right) \\
    &\quad - \tau\epsilon(\alpha) \left( \ket{\psi_+}\bra{\psi_+} - \ket{\psi_-}\bra{\psi_-} \right) \\
    &\quad + (1-\tau)\epsilon(\alpha) \left( -\ket{\psi_+}\bra{\psi_+} + \ket{\psi_-}\bra{\psi_-} \right) \\
    &= (E - \epsilon(\alpha)) \ket{\psi_+}\bra{\psi_+} + (E + \epsilon(\alpha)) \ket{\psi_-}\bra{\psi_-}
\end{aligned}
\end{equation}

Here, $E$ is the energy of the periodic unperturbed chain. Therefore along this path that avoids the point $\ta=\tb$ and connect the $\ta>\tb$ and $\ta<\tb$ regimes, the energy gap at the Brillouin zone boundary is constant.

The same analysis can be applied to the observed  gap closing at $k=0$ (even gaps). Here we have $\ket{\psi_+} = \ket{\psi_+^{1/2}}$, and  \(\ket{\psi_-} = -\ket{\psi_-^{1/2}}\). One eventually finds that, in this case, even gaps close at $\tau=1/2$.

\section{edge states in $AB^q$ sequences}
\label{app:ABn}

\begin{figure}[h]
\centering
\includegraphics[width=0.8\linewidth]{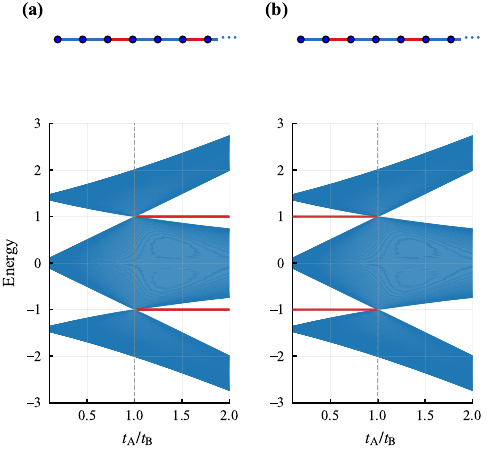} 
\caption{
Illustration of the peculiar edge states for OBC $AB^q$ chains, here in the simple case of $AB^2$ chains. We fix $\tb=1$ and let $\ta$ vary across the gap closing point, and show what happens when the consecutive $B$ bonds are placed on the left part of the first unit cell (left figure), and when the $A$ bond is shifted one step left (right figure). The spectrum shows $\ta$-independent edge states either in the $\ta<\tb$ side or in the $\ta>\tb$ side, as explained in the text.}
\label{fig:ABm spectrum}
\end{figure}

As said in the main text, for inversion-symmetric one-dimensional systems, a physical edge inherently breaks the inversion symmetry. Consequently the presence of edge states depends on the microscopic termination of the chain rather than on topological bulk properties. In this appendix, we show that, for $AB^q$ sequences, there are nevertheless interesting edge states, whose energies depend only on the $t_B$ hopping term, under specific open boundary conditions. This is shown in Fig.~\ref{fig:ABm spectrum}

We now provide a simple analytical explanation for this phenomenon.

Let us number the sites of the semi-infinite $AB^q$ chain as $(i,j)$, where $i= 1 \cdots \infty$ refers to the unit cell and $j=1\cdots n$ runs in the unit cell. We choose here a configuration such that $q$ $B$ bonds are on the left ($n-1$ first positions), and the bond $A$ on the right of the unit cell. The semi-infinite chain tight-binding basis now reads $\{\ket{i,j}\}$. What is found numerically is that the observed $\ta$-independent edge states occur precisely at the eigenvalues of an isolated small chain made of $q-1$ bonds with hopping terms $\tb$ (see Fig.~\ref{fig:ABm spectrum}-left for a $AB^2$ chain).

A general state reads (without caring about normalization) : 
$$
\ket{\Psi}= \sum_{i=1\ldots\infty} \sum_{j =1,\ldots,n} \psi_{i,j} \ket{i,j}
$$

We call $\ket{\phi}$ a normalized eigenstate of an isolated $q-1$-bond chain (which we call a short chain), with eigenvalue $E$, and introduce an  ansatz for the OBC chain eigenstate under  the form
\begin{equation}
\begin{split}
\psi_{i,j}&=\phi_{i,j} \qquad  j=1\ldots n-1 \\
\psi_{i,n}&=0.
\end{split}
\label{eq:ansatz}
\end{equation}

We have therefore positioned unnormalized copies of $\phi$, translated in each unit cell; these copies are not connected, being separated at sites with vanishing amplitude.

The condition under which this state can stand as an eigenvalue of the large OBC chain reads
\begin{equation}
\begin{split}
&H \ket{\psi} = E \ket{\psi}\\
&\ta \psi_{i+1,1} + \tb \psi_{i,n} = 0 \qquad \forall i,
\end{split}
\label{eq:ansatz}
\end{equation}

and therefore,we have \(\psi_{i,n} = - (\ta/\tb) \psi_{i+1,1}\).

Let us define the ratio
\begin{equation}
v_{i,i+1} \equiv \frac{\psi_{i,n}}{\psi_{i+1,n}} 
= \frac{\psi_{i,n}}{\psi_{i+1,1}} \frac{\psi_{i+1,1}}{\psi_{i+1,n}} 
= -\frac{\ta}{\tb} \frac{\psi_{i+1,1}}{\psi_{i+1,n}}.
\end{equation}

The short chains with uniform $\tb$ hopping terms are inversion symmetric, with symmetrical and antisymmetrical eigenstates. As a consequence, $|\psi_{i,1}|=|\psi_{i,n}|$ for any \(i\). Consequently, $v=|v_{i,i+1}|$ is a constant, equal to $|\ta/\tb|$ denoted as $v$. The wave function on all short chains then follows:
\begin{equation}
|\psi_i |= v^{-(i-1)}|\psi_1|.
\end{equation}

Whenever $\ta>\tb$, this gives an exponentially decaying wave function penetrating into the bulk, a hallmark of surface states, here with energy $E$. Since $E$ is an eigenvalue of the small chain with only $\tb$ hopping terms, this shows that, in that regime, the edge-state energy is independent of $\ta$.


Now let us consider the slightly modified OBC chain in Fig.~\ref{fig:ABm spectrum}-right, where the $A$ bond has been shifted one step on the left. We can repeat the same analysis as above, but now the condition on the site of vanishing amplitude has changed, and reads now
$$
\tb \psi_{i+1,1} + \ta \psi_{i,n} = 0 \qquad \forall i.
$$
As a consequence the ansatz state is indeed an eigenfunction of the OBC chain, but in the range $\ta<\tb$, as clearly seen in Fig.~\ref{fig:ABm spectrum}.

\section{Polarization in the presence of diagonal terms}
\label{app:diagonal term}

\begin{figure}[htbp]
    \centering
    \includegraphics[width=\linewidth]{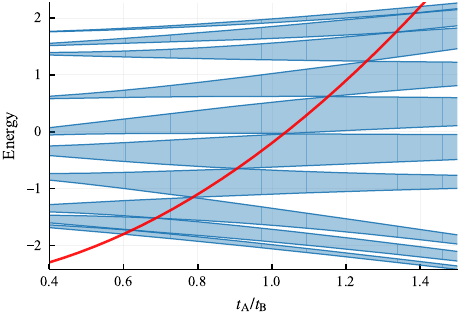}
    \caption{
    Spectrum as a function of $\ta/\tb$ for $\lambda=0.2$ and a CP chain of slope $\alpha=3/7$ ($n=10$). The red curve goes through all gap closures.}
    \label{fig:spectrum diagonal}
\end{figure}

In this work, we have mainly considered periodic CP chains with purely off-diagonal hopping terms. The essential ingredient is inversion symmetry, while the occurrence of TPTs follows an arithmetic criterion involving the number of sites $n$ in the unit cell and the number of occupied bands $M$. In this purely off-diagonal case, the transition takes place at the gap-closing point $\ta=\tb$.

It is natural to extend this analysis by introducing diagonal on-site terms that distinguish inequivalent local environments while preserving inversion symmetry. This can be achieved by assigning different on-site energies to sites connected to one $A$ and one $B$ bond, and to sites connected to two $B$ bonds. As shown in Ref.~\cite{SireMosseri1990}, the tight-binding spectrum of CP chains with $n$ sites per unit cell, including both diagonal terms $\pm\lambda/2$ and off-diagonal hoppings $(\ta,\tb)$, still consists of $n$ bands, but shows a different gap closing phenomenon: Gaps now close on a well-defined surface in the three-dimensional parameter space $(\lambda,\ta,\tb)$, while when $\lambda=0$, the gap closing condition is simply $t_A = t_B$.

\begin{figure}[htbp]
    \centering
    \includegraphics[width=\linewidth]{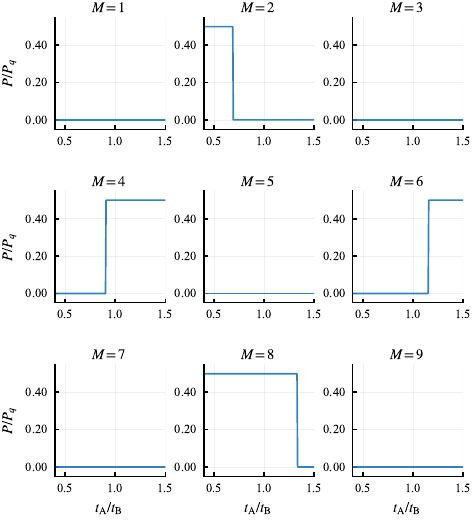}
    \caption{
    Polarization $P/P_q$ as a function of $\ta/\tb$ for $\lambda=0.2$ in a CP chain of slope $\alpha=3/7$ ($n=10$). Polarization jumps occur only for even values of $M$, exactly at the gap-closing points visible in Fig.~\ref{fig:spectrum diagonal}.}
    \label{fig:polarization diagonal}
\end{figure}

These chains being still inversion symmetric, our arithmetic criterion should therefore remain valid across these new gap closing points. Figure~\ref{fig:spectrum diagonal} shows the spectrum for fixed $\lambda=0.2$ as a function of $\ta/\tb$ for the chain of slope $\alpha=3/7$, with $n=10$. Several gap closings are clearly visible and occur at different values of $\ta/\tb$, along a red curve described in  Ref.~\cite{SireMosseri1990}. On this curve, the eigenstates are known exactly in terms of modified Bloch waves (waves with $E_\perp$ coordinates). 
We then compute the polarization to verify that the arithmetic criterion correctly predicts the occurrence of TPTs. As established in Sec.~\ref{sec:general even}, no TPT is expected for odd values of $M$, whereas TPTs occur for even $M$, since $\gcd(10,M)=2$ and $2$ does not divide $n/2=5$. The numerical results, shown in Fig.~\ref{fig:polarization diagonal}, are in complete agreement with this prediction.

\bibliography{Refs}

\end{document}